\documentclass[10pt]{article}
\usepackage{wws2024} 
\usepackage{url}
\usepackage{latexsym}
\usepackage{amsmath, amsthm, amsfonts}
\usepackage{algorithm, algorithmic}  
\usepackage{graphicx}
\usepackage{lipsum} 
\usepackage{fancyhdr}
\usepackage[headheight=1in,margin=1in]{geometry}
\usepackage{hyperref}

\title{ORES-Inspect: A technology probe for machine learning audits on enwiki}

\author{
Zachary Levonian \\
  Digital Harbor Foundation \\\And
Lauren Hagen \\
  University of Minnesota \\\And
Lu Li \\
  University of Pennsylvania \\\AND
Jada Lilleboe \\
  University of Minnesota \\\And
Solvejg Wastvedt \\
  University of Minnesota \\\And
Aaron Halfaker \\
  Microsoft Research \\\And
Loren Terveen \\
  University of Minnesota \\}

\begin{document}
\maketitle
\thispagestyle{fancy}

\begin{abstract}
Auditing the machine learning (ML) models used on Wikipedia is important for ensuring that vandalism-detection processes remain fair and effective.
However, conducting audits is challenging because stakeholders have diverse priorities and assembling evidence for a model's [in]efficacy is technically complex.
We designed an interface to enable editors to learn about and audit the performance of the ORES edit quality model.
ORES-Inspect\footnote{\href{https://ores-inspect.toolforge.org}{https://ores-inspect.toolforge.org}} is an open-source web tool and a provocative technology probe for researching how editors think about auditing the many ML models used on Wikipedia.
We describe the design of ORES-Inspect and our plans for further research with this system.
\end{abstract}


\section*{Introduction}

\begin{figure*}[htp]
\begin{center}
\centerline{\includegraphics[width=\textwidth]{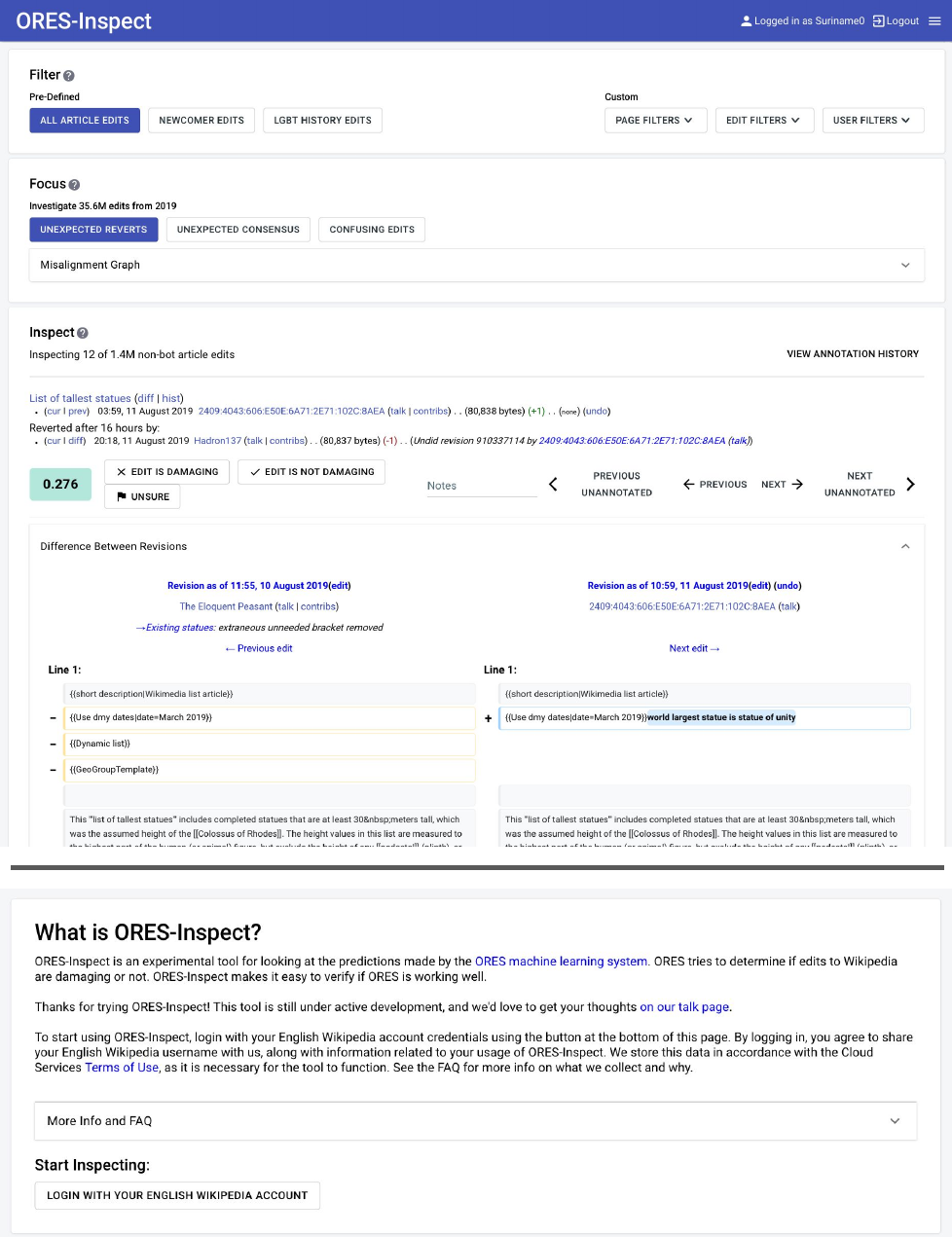}}
\caption{The ORES-Inspect interface and login page, as accessible via Toolforge.}
\label{fig:interface}
\end{center}
\end{figure*}

ORES is a widely-used service for building and hosting machine learning models requested by the Wikipedia community \cite{halfaker_ores_2020}.
Of particular relevance is the edit quality model, which makes predictions about the quality of individual Wikipedia edits and is used in other systems for vandalism detection and removal.
ORES' edit quality predictions directly influence the likelihood of an edit being reverted \cite{teblunthuis_effects_2020}.
This impact is a notable success for community-centered and participatory machine learning processes: ORES is hosting an increasing number of models.\footnote{ORES is being \href{https://wikitech.wikimedia.org/wiki/ORES}{replaced with LiftWing}, but this work is applicable to any revscoring model.}

A key challenge for ORES and other machine learning services is that it is hard to determine if a model is consistently producing reasonable outputs.
In other words, it is hard to \textit{audit} these models.
There are many barriers to auditing complex machine learning systems like ORES: (a) identifying a relevant sample of incorrect predictions, (b) determining if those incorrect predictions represent a pattern of undesired behavior (a ``bug''), and (c) convincing system designers to fix the undesired behavior.
To address those barriers, we are building ORES-Inspect, an open-source\footnote{\href{https://github.com/levon003/wiki-ores-feedback}{https://github.com/levon003/wiki-ores-feedback}} tool to audit the behavior of the ORES edit quality model for English Wikipedia.

In the consensus-driven Wikipedia context, the developers of ML-driven systems like ORES are enthusiastic about receiving community input on problems or potential areas for improvement. 
Thus, the key design objective for ORES-Inspect is to address problems (a) and (b) by making it easy to identify high-quality quantitative evidence of the ORES edit quality model's behaviors. 
We are developing ORES-Inspect as a ``technology probe'' to reflect on the process of conducting ML audits in the Wikipedia context by highlighting the benefits and challenges of collecting quantitative evidence of system bugs \cite{hutchinson_technology_2003}.

Functionally, ORES-Inspect is a labeling interface for individual Wikipedia edits.
The key intuition is that any Wikipedia user may be interested in auditing a system like ORES, but different auditors will have different priorities (e.g. are new editors unfairly targeted, is vandalism on stubs missed more often than on larger articles, etc.).
For that reason, the process of auditing is the process of quantifying one's intuitions and identifying evidence that a single misclassification represents a pattern that should be changed.
Therefore, we designed ORES-Inspect as a provocation: it is designed to educate editors about how ML models can be audited and how to translate intuitions into high-quality evidence.

To fulfill this educational objective and to make auditing tractable for users, we designed the interface (Figure~\ref{fig:interface}) around four phases of activity.
We will describe our design decisions, the data, and our future analysis plans in the remainder of this extended abstract, but we conclude this introduction with the verbatim contents of the info panel shown to ORES-Inspect users on first login:

\textbf{ORES finds vandalism.}
ORES is a machine learning model that gives every edit on Wikipedia a score from 0 (least likely to be damaging) to 1 (most likely to be damaging). Score predictions are used to highlight the Recent Changes feed and in other places to find and revert vandalism. ORES-Inspect helps you audit ORES by looking at score predictions and determining if they are correct.
Audit ORES in four steps:


\begin{enumerate}
    \item \textbf{Filter:}
Choose which edits to look at. ORES-Inspect shows you all human edits on mainspace articles by default, but you can filter down to look only at edits on particular pages (such as pages related to LGBT history) or from particular editors (such as newcomers).

Or, use the filter controls to choose something else entirely, like bot edits on Talk pages!
\item \textbf{Focus:}
When an edit is damaging, it is usually reverted by the editor community. ORES-Inspect helps you focus on cases where the community behavior disagrees with the ORES prediction.

If you choose to look at Unexpected Reverts, you're looking at edits that ORES thinks are non-damaging...\ but that the community reverted anyway.

If you choose to look at Unexpected Consensus, you're looking at edits that ORES thinks are damaging... but that the community didn't revert.

\item \textbf{Inspect:}
Look at individual edits and label them as damaging (``I would revert this.'') or not damaging. See if you can find a pattern of errors in ORES' predictions.

\item \textbf{Discuss:}
View a summary of your edit labels by clicking ``View Annotation History''. How often did ORES misclassify the edits you looked at?

You can discuss your results with the ORES developers. If you change the filters, you can compare two groups of edits to identify bias (``Are newcomers' edits misclassified more often than experienced editors'?'')
\end{enumerate}

\section*{Implementation \& Data}

ORES-Inspect is a React and Python app hosted on Toolforge.
Auditors use filters to focus their attention on specific properties of articles (namespace, category, size), of edits (size, marked as minor), or of users (registration status, bot status).
ORES-Inspect is based on the 35.6 million non-bot enwiki edits in 2019 and the corresponding prediction made by ORES at the time of the edit,\footnote{Historical ORES predictions were only available until the end of 2019.} but auditors focus on only those revisions that have already received attention by the community: Unexpected Consensus edits are predicted to be damaging by ORES but \textit{were not} reverted within 1 year, while Unexpected Reverts were predicted to be non-damaging by ORES but \textit{were} reverted.
By focusing on these two categories, we focus on identifying false positives and false negatives respectively with a much higher precision than random sampling of revisions.
By then inspecting and labeling specific revisions as damaging or not damaging, auditors create quantitative estimates of the prevalance of false positives and/or false negatives for a specific subset of pages, edits, or editors.

\section*{Discussion \& Future Work}

Other research-driven interfaces for working with ORES include Wikibench for curating and discussing training data \cite{kuo_wikibench_2024} and ORES Explorer for exploring fairness trade-offs induced by model thresholding decisions \cite{ye_wikipedia_2021}.
We focus on auditing models that are already in use with an emphasis on building quantitative evidence of ML system bugs.
In our experience as Wikipedia editors, we observe that most feedback on ML systems happens on the basis of a single bad prediction noticed while focused on other editing work. ORES-Inspect aims to be a tool for turning those singletons into rigorous and useful audits, and we have already found the tool helpful for reflecting on how one's opinions on edit quality might diverge from consensus. 
As we continue to develop ORES-Inspect, we intend to conduct interviews with editors and share the results of audits conducted with the tool, aiming to generate discussion on how and when ML model efficacy should be evaluated by editors.




\section*{Acknowledgements}

We would like to thank Haiyi Zhu, Phyllis Gan, and Isaac Johnson for their contributions and the Wiki Workshop reviewers for useful feedback.
This work was partially supported by the National Science Foundation under Grant No.\ 1908688.

\bibliographystyle{wws2024} 
\bibliography{wiki_misclass}


\end{document}